\documentclass[11pt]{article}
\usepackage[protrusion=true,expansion=true]{microtype}
\usepackage{amsmath,amsfonts, braket}
\usepackage{tikz}
\usetikzlibrary{trees,er,snakes,shapes,mindmap}
\usepackage{titlesec}
\titleformat{\section}
 {\normalfont\fontfamily{put}\fontsize{12pt}{16pt}\bfseries\color{black}}
{\thesection}{1em}{}
\def \beq  {\begin{equation}}
\def \eeq  {\end{equation}}
\def \beqar {\begin{eqnarray}}
\def \eeqar {\end{eqnarray}}
\allowdisplaybreaks
\def\sqr#1#2{{\vcenter{\vbox{\hrule height.#2pt
\hbox{\vrule width.#2pt height#1pt \kern#1pt
\vrule width.#2pt}\hrule height.#2pt}}}}

\def\S {{\cal S}}
\def\la {{\langle}}
\def\ra {{\rangle}}
\def\vx {{\vec x}}
\def\vy {{\vec y}}
\def\vk {{\vec k}}

\def\Tr {{\rm Tr}}

\def\vk {\vec{k}}
\def\vx {{\vec x}}

\def\vy{\vec{y}}

\def\del {\partial}

\def\bz {{\bar{z}}}

\def\A {{\cal A}}
\def\B {{\cal B}}

\def\D {{\cal D}}

\def\H {{\cal H}}
\def\I{{\cal I}}

\def\M{{\cal M}}
\def\O {{\cal O}}
\def\P {{\cal P}}

\def\half{\textstyle{1\over 2}}

\mathchardef\mhyphen="2D
\begin{document}
\fontfamily{put}\fontsize{11pt}{16pt}\selectfont
\def \CMP {{Commun. Math. Phys.}}
\def \PRL {{Phys. Rev. Lett.}}
\def \PL {{Phys. Lett.}}
\def \NPBProc {{Nucl. Phys. B (Proc. Suppl.)}}
\def \NP {{Nucl. Phys.}}
\def \RMP {{Rev. Mod. Phys.}}
\def \JGP {{J. Geom. Phys.}}
\def \CQG {{Class. Quant. Grav.}}
\def \MPL {{Mod. Phys. Lett.}}
\def \IJMP {{ Int. J. Mod. Phys.}}
\def \JHEP {{JHEP}}
\def \PR {{Phys. Rev.}}
\def \JMP {{J. Math. Phys.}}
\def \GRG{{Gen. Rel. Grav.}}
\begin{titlepage}
\null\vspace{-62pt} \pagestyle{empty}
\begin{center}
\vspace{1.3truein} {\Large\bfseries
Entanglement for Quantum Hall states and a Generalized}\\
\vskip .15in
{\Large\bfseries Chern-Simons Form}\\
\vskip .5in
{\Large\bfseries ~}\\
 {\large\sc V.P. Nair}\\
\vskip .2in
{\itshape Physics Department,
City College of the CUNY\\
New York, NY 10031}\\
 \vskip .1in
\begin{tabular}{r l}
E-mail:&\!\!\!{\fontfamily{cmtt}\fontsize{11pt}{15pt}\selectfont vpnair@ccny.cuny.edu}\\
\end{tabular}
\vskip 1.5in
\centerline{\large\bf Abstract}
\end{center}
We analyze some features of the entanglement entropy for an integer quantum Hall state
($\nu =1 $)
in comparison with ideas from relativistic field theory and noncommutative geometry.
The spectrum of the modular operator, for a restricted class of states,
 is shown to be similar to the case of field theory or a type ${\rm III}_1$ von Neumann algebra.
We present arguments that the main part of the dependence of the entanglement entropy on background fields and geometric data such
 as the spin connection is given by a generalized Chern-Simons form.
 Implications of this result for bringing together ideas of noncommutative geometry, entropy and gravity
 are briefly commented upon.

\end{titlepage}
\pagestyle{plain} \setcounter{page}{2}
\section{Introduction}
The idea that there is some deep connection between entropy and gravity is 
by now well-known and
well-accepted \cite{TJ}-\cite{verlinde}. Also the Reeh-Schlieder and Connes-Stormer theorems \cite{CS}, coupled with the observation that
the algebra of local observables should be a type ${\rm III}_1$ von Neumann algebra,
tell us  that entanglement is an integral part of relativistic 
quantum field theory \cite{vN}-\cite{witten}.
In noncommutative geometry, one attributes degrees of freedom to
space itself via its description by the states of a suitable Hilbert space \cite{NC}.
Putting these three observations together, a question which naturally arises is
whether we can calculate an entanglement entropy
between states defining the spatial geometry
in the noncommutative scenario and relate it to gravity.
This is the subject we explore in this paper.

A simple working model for noncommutative geometry is given by the quantum Hall system.
If we consider a K\"ahler manifold $\M$, we can choose a background magnetic field
which is proportional to the K\"ahler two-form. The lowest Landau level (LLL) is obtained by
quantizing $\M$ with this multiple of the K\"ahler form as the symplectic structure.
We get a Hilbert space $\H$ which is spanned by holomorphic wave functions
and which can be used as the model for the noncommutative version
of $\M$ \cite{KNrev}, \cite{KN1}.
Although we phrased this in terms of a Landau-Hall problem,
the states of the LLL can be viewed as holomorphic 
sections of a power of the canonical line bundle, so the tie-in to the
physical situation of the Hall effect is useful but not essential.
On the other hand, what we do in this paper can also be viewed
more narrowly as an interesting view on some entanglement issues for the Hall system, 
ignoring the larger perspective of gravity and noncommutative geometry.

We consider a completely filled LLL and a surface
(of co-dimension 1) separating $\M$ into two regions. 
It is then possible to define an algebra of local observables
and reduced density matrices. 
Regarding the
entanglement between the states in the two regions,
we consider the spectrum of the modular operator and show that
it is basically ${\mathbb R}_+$. 
If we consider only the fully filled LLL, which is what is relevant in modeling
noncommutative geometry,
the only freedom in the reduced density matrices is due to a change of the
separating surface in $\M$.
For all such cases, we will see that the spectrum of the modular operator is
${\mathbb R}_+$, as the number of states tends to infinity.
In relativistic quantum field theory, the
algebra of local observables is expected to be a type ${\rm III}_1$
von Neumann algebra. This means that the intersection of
the spectra of the modular operators over all states (or density matrices)
should be 
${\mathbb R}_+$ \cite{vN}, \cite{HM}. In the present problem, we do not exactly
have this result as we are not
considering all possible density matrices.
For the fully filled LLL, as mentioned above,
only a smaller class of density matrices is meaningful.
Over this
set of reduced density matrices, we do obtain the same spectrum, namely, ${\mathbb R}_+$.

The second part of our analysis focuses on the changes in the entanglement entropy
as the background fields are varied.
We can consider fluctuations in the magnetic field as well as changes in
the background geometry due to gravitational fluctuations.
We show that the
entanglement entropy, as a function of these background fields,
is proportional to a Chern-Simons action.
It should be emphasized that we are not discussing the effective action for
the Hall states; it would not be deemed surprising that the latter
is a Chern-Simons theory. To highlight the nature of our result, consider the fact that
Einstein gravity in 2+1 dimensions is described by a 
Chern-Simons theory \cite{3dgrav}. There have also been investigations of
higher dimensional Chern-Simons gravities recently \cite{zanelli}.
In these theories, the gravitational field equations
are the extremization of the CS action.
Our argument shows that they may be related to the extremization of the
entanglement entropy between states corresponding to the degrees of freedom of
space itself.

In section 2, we recall a few relevant results from relativistic quantum field theory.
The spectrum of the modular operator for Landau-Hall states on $S^2$
(or fuzzy version of $S^2$) is considered in section 3.
The generalization to ${\mathbb{CP}^k}$, $k>1$, is done in section 4.
The background field dependence of the entanglement entropy is discussed in
section 5. The paper concludes with a short discussion and two appendixes with 
explicit details of 
some of the relevant calculations.

\section{Observations from Field Theory}

In this section, we collect a few known observations about relativistic quantum field theory
which can serve as points of comparison for our analysis for Hall states.

A key property of relativistic QFT is that local observables commute at spacelike separations,
\beq
[ \phi (x), \phi (y) ] = 0, \hskip .2in
(x-y)^2 < 0
\label{qft1}
\eeq
Here $\phi(x)$ are not necessarily fundamental or elementary fields.
In some neighborhood $\O$ of spacetime, we can define a local algebra of observables,
denoted by $\A (\O )$,
defined by bounded operators of the form
\beq
\phi (f) = \int f (x)\, \phi (x)
\label{qft2}
\eeq
where the support of $f(x)$ is contained in $\O$.
$\A (\O) $ forms a subalgebra of  ${\cal B} (\H)$, the set of bounded operators
on the Hilbert space $\H$.
It is unital in the sense that it includes the identity and is a $*$-algebra since it
inherits an involution corresponding to the adjoint operation.
If $\O'$ is the causal complement of $\O$, then (\ref{qft1}) translates as
\beq
[ \A (\O), \A(\O')] = 0
\label{qft3}
\eeq
This tells us that $\A (\O')$ is contained in the commutant $\A'$ of $\A (\O)$, namely the set of all
operators which commute with $\A (\O)$.
Following Haag, we take $\A (\O') = \A(\O)'$, a statement which is known as Haag duality.
We also assume that $\A (\O) = \A(O)''$.
A unital $*$-subalgebra $\A$ of $\B(\H)$ 
which has the property $\A = \A''$ is a
von Neumann algebra. (There are other definitions based on operator topology, but this is
the simplest for our purpose.)
Thus, we can treat $\A (\O)$ and $\A (\O') \, (= \A (\O)' \,)$ as von Neumann algebras.
In what follows, we will consider 
fields at a given time. Strictly speaking, the definition of local operators will need
point-spitting in time, but this refinement will not be important for most of the following
discussion.

While one can define an algebra of local observables, the Hilbert space of states does not
factorize into Hilbert subspaces defined locally. This statement is the result of some deep
theorems, but a simple illustrative example which highlights this
feature is 
obtained in terms of local single-particle states.
Consider defining
 ``local" one-particle states of the form
\beq
\ket{f} = \int d^3x \, f(x) \, \psi^\dagger (x) \ket{0},
\hskip .2in
\ket{h} = \int d^3x \, h(x) \, \psi^\dagger (x) \ket{0}
\label{qft4}
\eeq
where the functions $f(x)$ and
$h(x)$ have supports in disjoint regions of space and $\psi^\dagger$ denotes the
negative frequency (creation) part of an elementary field operator, 
which may be taken, for the present purpose, as a free field for simplicity.
The overlap of these states is given by
\beq
\braket{f|h} = \int d^3x d^3y\, {d^3 k \over (2 \pi)^3} {1\over 2 \omega_k} 
e^{-i\vk\cdot (\vx - \vy )} \, f(x) \, h(y)
\label{qft5}
\eeq
where $\omega_k = \sqrt{k^2 + m^2}$. (We consider particles of mass $m$.)
The factor $1/(2\omega_k)$, which is characteristic of the
relativistic theory, plays a crucial role. In the nonrelativistic case where $\omega_k \approx m$,
this factor is a constant, independent of $k$, and
the integration over $k$ gives a $\delta$-function and hence the overlap integral
is zero since $f$ and $h$ have no overlap for their supports.
But in the relativistic theory, we see that this overlap is nonzero, rendering void any attempt
to define local one-particle states.
Entanglement thus becomes a characteristic feature of relativistic field theory.

Another important feature is embodied in the Reeh-Schlieder theorem which 
tells us that the local algebra $\A (\O)$ is sufficient to generate a dense set of states
on the Hilbert space of the theory by their action, say, on the vacuum state \cite{witten}. 
In other words, the vacuum state $\ket{0}$ is a cyclic vector for the algebra
$\A (\O)$. For such a state, if $A' \ket{0} = 0$ for
$A' \in \A'$, then
\beq
 0 = A\, A' \ket{0} = A' \, A \ket{0}
 \label{qft5a}
 \eeq
 where we use the fact that $A$ and $A'$ commute. Since $A \ket{0}$ generates a dense set
 of states, considering all $A\in \A$, we see that $A'$ should vanish on a dense set of 
 states, hence $A' = 0$. A vector $\ket{\Psi}$ is said to be a separating vector
if $A' \ket{\Psi} = 0$ implies
 $A' = 0$. Thus a cyclic vector for $\A$ is separating for
 $\A'$ and vice versa. Similar arguments apply starting from $\A'$, so we have the result that
 the vacuum $\ket{0}$ is cyclic and separating for both $\A(\O)$ and $\A (\O)'$\footnote{
 A related statement or corollary is that there is no bounded local operator which annihilates
the vacuum state, a statement which is useful for proving Coleman's theorem on realizations of symmetry.}.
This is the required premise for the Tomita-Takesaki theorem.
Towards the statement of the theorem, let
$\ket{\Psi}$ is a cyclic and separating vector for the von Neumann algebra $\A$.
(We will use a general state $\ket{\Psi}$ for many of the statements here, 
although specializing to the
vacuum state $\ket{0}$ is most pertinent to the general field theory analysis.)
We can then define an antilinear map $S_\Psi$ whose action is given by
\beq
S_\Psi ( A \ket{\Psi}) = A^\dagger \, \ket{\Psi}
\label{qft6}
\eeq
Evidently $S_\Psi^2 = 1$. Since $S_\Psi$ is antilinear, conjugation is defined by
(see \cite{witten}),
\beq
\braket{ \alpha | S \beta } = \overline{\braket{S^\dagger \alpha| \beta}} =
\braket{\beta| S^\dagger \alpha}
\label{qft6a}
\eeq
Thus if we define $F A'\ket{\Psi} = A'^\dagger \ket{\Psi}$, 
\beqar
\braket{A' \Psi | S A \Psi} &=& \braket{A' \Psi | A^\dagger \Psi} 
= \braket{ AA' \Psi | \Psi} = \braket{A'  A\Psi | \Psi} 
= \braket{A \Psi | A'^\dagger \Psi} \nonumber\\
&=& \braket{A \Psi | F A' \Psi} 
\label{qft6b}
\eeqar
Taking $\ket{\alpha} = \ket{A'\Psi}$, $\ket{\beta} = \ket{A \Psi}$, we see from
(\ref{qft6a}) that $F= S^\dagger$. Thus $S^\dagger$ acts on $\A'$ as $S$ acts on
$\A$. Going back to (\ref{qft6}), the states $A \ket{\Psi}$ and $A^\dagger \ket{\Psi}$ 
do not have the same norm in general, so $S_\Psi$ is not unimodular. We can separate out
a unimodular part $J$ via the polar decomposition $S = J \, \Delta^{{1\over 2}}$.
where $J$ is a unimodular antilinear operator and $\Delta$ is self-adjoint.
$J$ is referred to as the modular conjugation and $\Delta$ is the Tomita
modular operator.
The latter can also be defined by
\beq
\Delta_\Psi = S^\dagger_\Psi S_\Psi
\label{qft6c}
\eeq
This operator depends on the choice of the state $\ket{\Psi}$.
Among other useful properties of $S$, $S^\dagger$, we can easily  verify
that $J^\dagger = J$ and
\beqar
J^2 = J^\dagger J = 1, &\hskip .2in&J \Delta^{1\over 2} J = \Delta^{-{1\over 2}}\nonumber\\
S^\dagger_\Psi = \Delta^{\half}_\Psi \, J^\dagger, &\hskip .2in&
S_\Psi S^\dagger_\Psi = \Delta_\Psi^{-1}
\label{qft6ca}
\eeqar

The Tomita-Takesaki theorem is the statement that
given a von Neumann algebra $\A$ and a cyclic and separating vector
$\ket{\Psi}$, with $S$ and $S^\dagger$ as defined above,
\beqar
J \, \A \, J &=& \A' \nonumber\\
\Delta^{it} \, \A \, \Delta^{-it} &=& \A, \hskip .2in
{\rm for~all} ~~t \in \mathbb{R}
\label{qft6d}
\eeqar
The first statement relates $\A$ and $\A'$, while the second identifies a one-parameter family of
automorphisms of $\A$ which may be viewed as time-evolution.
The proof of the theorem is very involved, we do not discuss it,
but to see how such statements could arise, notice that, if $A, \, B \in \A$,
\beqar
S B S \, A \ket{\Psi} &=& S B \, A^\dagger \ket{\Psi} = A B^\dagger \ket{\Psi}\nonumber\\
A \, S B S \ket{\Psi} &=&  A S B \ket{\Psi} = A B^\dagger \ket{\Psi}
\label{qft6e}
\eeqar
Thus $SBS$ is contained in $\A'$. Reducing this to the unimodular part of
the action of $S$, we see how the first part of the theorem could arise.

Using the polar decomposition, (\ref{qft6}) gives $J \Delta^{{1\over 2}} A \ket{\Psi}
= A^\dagger \ket{\Psi}$. Thus $\Delta^{{1\over 2}} A \ket{\Psi}$ and $A^\dagger \ket{\Psi}$
should have the same norm.
Taking $\ket{\Psi}$ to be the vacuum $\ket{0}$, since
$A$ is in the algebra of local observables in $\O$, we can define
the unimodularity condition for $J$ as
\beq
\Tr ( \rho_\O A^\dagger \Delta A ) = \Tr ( \rho_{\O} A A^\dagger )
\label{qft6f}
\eeq
where $\rho_\O$ is the reduced density matrix starting from
$\ket{0}$. 
 (We may view the vacuum in terms of its wave function as a functional of
 the fields and we can integrate out the part of the fields corresponding to
 $\O'$ to obtain this. How exactly this is done is not important for now.)
 There is a similar equality for operators in $\A'$ with a reduced density
matrix $\rho'_{\O'}$.
Notice that the relation (\ref{qft6f}) and the corresponding one for operators
in $\A'$ are obtained if we define the action of $\Delta$ by
\beqar
\Delta\, A = \rho_\O \rho'^{-1}_{\O'} \, A\, \rho'_{\O'} \rho^{-1}_\O\nonumber\\
\Delta\, A' = \rho_\O \rho'^{-1}_{\O'} \, A'\, \rho'_{\O'} \rho^{-1}_\O
\label{qft6g}
\eeqar
Since the vacuum state is a separating vector, $\rho_{\O}^{-1}$ and $\rho_{\O'}^{-1}$
do exist and these
formulae are well defined.
Introducing states $\ket{k, \tilde{k}}$ which form a basis for
 $\H\otimes \H$, we can represent the
operators $A$ and $A'$ in the form\footnote{We use discrete labels and summation signs
to give the general tenor of the results and
to write expressions in a form suitable for later sections. An appropriate limit
will be needed for the continuum field theory.}
\beq
A = \sum_{k, \tilde{k}} \ket{k, \tilde{k}} \, (A_{kl} \,\delta_{\tilde{k} \tilde{l}} )\,
\bra{l, \tilde{l}} ,\hskip .2in
A' = \sum_{k, \tilde{k}} \ket{k, \tilde{k}} \, (\delta_{kl}\,A'_{\tilde{k} \tilde{l}} )\,
\bra{l, \tilde{l}}
\label{qft6h}
\eeq
The action of $\Delta$ as in (\ref{qft6g}) can then be written as
\beq
\Delta \, \ket{k, \tilde{k}} = \sum_{l, \tilde{l}} (\rho_\O)_{kl} \, (\rho'^{-1}_{\O'})_{\tilde{k} \tilde{l}}\,
\ket{l, \tilde{l}}
\label{qft6i}
\eeq
A more convenient notation is to represent $\ket{k, \tilde{k}}$ as
$\Phi = \ket{k} \bra{\tilde{k}}$, so that the action of
$\Delta$ can be represented as \cite{witten}
\beqar
\Delta_\Psi \, \Phi &=& \sum_{l, \tilde{l}}  
(\rho_\O)_{kl} \, (\rho'^{-1}_{\O'})_{\tilde{k} \tilde{l}}\, \ket{k} \bra{\tilde{k}}
= \sum_{l, \tilde{l}}  (\rho_\O)_{kl} \,\ket{k} \bra{\tilde{k}}\, 
[(\rho'^{T}_{\O'})^{-1}]_{\tilde{l} \tilde{k}}\nonumber\\
&=& \rho_\O \, \Phi \, (\rho'^T_{\O'})^{-1}
\label{qft6j}
\eeqar
A similar equation holds for 
a more general state $\Phi = \sum c_{k, \tilde{k}} \ket{k, \tilde{k} }
= \sum c_{k, \tilde{k}} \ket{k}\bra{\tilde{k} }$.
It should be emphasized that the dependence of $\Delta_\Psi$ on the state
$\ket{\Psi} = \ket{0}$ is carried by the reduced density matrices, while $\Phi$
denotes another possible state for the algebras.

From the point of view of general analyses of
von Neumann algebras, the
importance of the modular operator $\Delta$ is that its spectrum can be used to
classify such algebras \cite{vN}, \cite{HM}.
The von Neumann algebra of local operators in relativistic quantum field theory
is expected to be of the hyperfinite Type III$_1$. This is characterized by
the property that
\beq
\bigcap_{\Psi } {\rm Spec} (\Delta_\Psi ) = {\mathbb R}_+
\label{qft7}
\eeq
In other words, the intersection of the spectra over all choices of the state
$\ket{\Psi}$
is ${\mathbb R}_+$. The spectrum itself is defined by the
action of $\Delta_\Psi$ on a general state $\Phi$ as in
(\ref{qft6j}). 
A consequence of the algebra being of the hyperfinite Type III$_1$ is 
that any state can be brought arbitrarily close to any other state
by unitary transformations defined separately in $\A (\O)$ and $\A (\O')$.
This is the essence of the Connes-Stormer theorem and implies that 
almost all states are entangled \cite{vN}, \cite{HM}.

In the light of these facts about relativistic quantum field theory,
we first consider the natural question of
whether,
for a quantum Hall state on a manifold divided into two regions, the spectrum of 
the modular operator is ${\mathbb R}_+$. We show that this is indeed the 
case in a limited sense. This is discussed in the next two sections.
We shall then take up the question of how the entanglement entropy
depends on the backgrounds gauge fields and the spin connection.

\section{The spectrum of the modular operator for the \,$\nu = 1$ Hall state}

We start by considering the quantum Hall state on the two-sphere $S^2$
where the lowest Landau level is fully occupied, i.e., the $\nu =1$ state.
The fermion field operators can be expanded as
\beq
\psi = \sum_s a_s \, u_s(x) + \sum_\alpha a_\alpha \, U_\alpha (x)
\label{vN1}
\eeq
where $u_s(x)$ are the single particle wave functions for the lowest Landau level
(LLL). $U_\alpha$ denote the higher Landau level wave functions, which
will not be very important for what follows.
The fully occupied state LLL can thus be specified as
\beq
\ket{\nu = 1} = a_0^\dagger \ a_1^\dagger \ \cdots a_n^\dagger \ket{0}
\label{vN2}
\eeq
where $n+1$ denotes the number of states which constitute the LLL. 
For $S^2$, $n = 2 B r^2$ where
$B$ is the radial magnetic field of a monopole at the origin in the standard embedding
of $S^2$ in ${\mathbb R}^3$.
From (\ref{vN1}), the annihilation and creation operators for the
LLL may be expressed as
\beq
a_s = \int d \mu \, u^*_s \psi , \hskip .2in
a^\dagger_s = \int d\mu\, u_s \psi^\dagger
\label{vN3}
\eeq
We can parametrize the sphere in terms of complex coordinates $z$, $\bz$, corresponding to 
the stereographic projection of $S^2$ onto the plane. The wave functions are then given by
\beq
u_s (x) = {1\over \sqrt{\pi}} \sqrt{ { \Gamma (n+2) \over s! \,\Gamma (n-s +1)}}
\,\, {z^s \over (1+ \bz z)^{n/2}}
\label{vN4}
\eeq

We want to separate the sphere into two regions, say, the northern hemisphere and the southern hemisphere. The equator, which is the dividing line, corresponds to
$\vert z\vert = 1$ in the coordinates we are using.
We thus define
\beqar
b_s = {1\over \sqrt{\lambda_s}} \int_0^{\vert z\vert =1} d\mu \, u^*_s \psi , 
&\hskip .1in&
b^\dagger_s = {1\over \sqrt{\lambda_s}} \int_0^{\vert z\vert =1} d\mu \, u_s \psi ^\dagger
\nonumber\\
c_s = {1\over \sqrt{1-\lambda_s}} \int_{\vert z\vert =1}^\infty d\mu \, u^*_s \psi , 
&\hskip .1in&
c^\dagger_s = {1\over \sqrt{1-\lambda_s}} \int_{\vert z\vert =1}^\infty d\mu \, u_s \psi ^\dagger
\label{vN5}
\eeqar
where
\beq
\lambda_s = \int_0^{\vert z \vert = 1} u^*_s u_s
\label{vN6}
\eeq
In terms of these operators
\beqar
a_s &=& \sqrt{\lambda_s}\, b_s + \sqrt{1-\lambda_s} \, c_s \nonumber\\
a^\dagger_s &=& \sqrt{\lambda_s}\, b^\dagger_s + \sqrt{1-\lambda_s} \, c^\dagger_s 
\label{vN7}
\eeqar
The operators $\{ b_s , b^\dagger_s\}$ and $\{ c_s, c^\dagger_s\}$ form two mutually commuting
fermions algebras, obeying
\begin{align}
\{b_s , b_r \} = \{ b^\dagger_s, b^\dagger_r\} = ~& 0  =
\{c_s , c_r \} = \{ c^\dagger_s, c^\dagger_r\}\nonumber\\
\{b_s, b^\dagger_r \} = ~&\delta_{rs} = \{c_s, c^\dagger_r \} 
\label{vN8}
\end{align}
The second set of commutation rules requires the definition of
 the normalization factor of $\sqrt{\lambda_s} $,
$\sqrt{1-\lambda_s}$ in (\ref{vN5}).
In verifying (\ref{vN8}), we also assume that the angular integrations suffice to
make the integral vanish for $r\neq s$. This is indeed the case and will be important for
the higher dimensional generalization.

A short parenthetical remark may be useful before we go on. If we consider functions which have
support only in the region $\vert z\vert < 1$, the lowest Landau level wave functions $\{ u_s\}$
are not an adequate basis for a mode expansion of such functions. One can get a complete basis
 by including the higher Landau levels as well. 
 This is also clear from using the full mode expansion (\ref{vN1}) for $\psi$ and
 $\psi^\dagger$ in  (\ref{vN5}). We then see that the operator
 expressions for $b_s, b^\dagger_s$ and
 $c_s, c^\dagger_s$ will also involve $a_\alpha, a^\dagger_\alpha$ where the subscript
 $\alpha$ refers to the
 higher LLs.

Returning to the main chain of reasoning, a state vector for
one fermion occupying the state corresponding to $u_s$ is given by
\beq
\ket{s} = a^\dagger_s \ket{0} = \sqrt{\lambda_s}\,
b^\dagger_s \ket{0} + \sqrt{1-\lambda_s} \, c^\dagger_s \ket{0}
\label{vN9}
\eeq
This defines a way of splitting the state in terms of degrees of freedom corresponding to
the inside region $\vert z \vert <1$ and the outside region
$\vert z \vert > 1$.
In fact, we can consider local observables which correspond to independent unitary transformations
of the $b_s$'s and the $c_s$'s.
Thus let $\A$ denote the set of all unitary transformations $U_{sr}$ on $b_s$
(and $b^\dagger_s$), of the form
$b_s \rightarrow U_{sr} b_r$, and ${\tilde \A}$ denote the
the set of unitary transformations $V_{sr} $ of the form
$c_s \rightarrow V_{sr} c_r$. These may be interpreted as the algebra of observables
for the region inside (i.e., $\A$) and the region outside (i.e., $\A' = \tilde{\A}$), respectively.
Evidently, these form two mutually commuting algebras
which are copies of $U(n+1)$,
\beq
[ \A, {\tilde \A}] = 0
\label{vN10}
\eeq
The state (\ref{vN9}) is, of course, entangled, 
since $\lambda_s \neq 0, 1$ in general.
Since the only operator which commutes with all of $U_{sr}$ is the identity
and similarly for $V_{sr}$, we see that $\A$ is the commutant of ${\tilde \A}$ and vice versa.
Thus we have two von Neumann algebras, which become infinite dimensional as
we take $N \rightarrow \infty$.
The state corresponding to
(\ref{vN2}), expressed in terms of the algebra as a density matrix,
can be written, upon using (\ref{vN9}), as
\beq
\rho = \prod_{\otimes_s} \Bigl[ \lambda_s~ b_s^\dagger \ket{0}\! \bra{0} b_s
+ (1 - \lambda_s )~c^\dagger_s \ket{0}\! \bra{0} c_s
+\sqrt{\lambda_s (1- \lambda_s)}~
\left( b^\dagger_s \ket{0}\! \bra{0} c_s + c^\dagger_s \ket{0} \!\bra{0} b_s \right)\Bigr]
\label{vN11}
\eeq
Notice that the state $c^\dagger_s \ket{0}$ has ``one particle of the $c$-type" although the occupation number for the $b$-type is zero and vice versa.
We can now trace over the $c$-states to get a reduced density matrix for the
$b$-type, and similarly for the $c$-type. These are given by
\beqar
\rho_b &=&\prod_{\otimes_s} \Bigl[ \lambda_s~b^\dagger_s \ket{0}\!\bra{0} b_s + (1- \lambda_s )  ~ \ket{0}\! \bra{0}\, \Bigr] \equiv \prod_{\otimes_s} (\rho_b)_s
\nonumber\\
\rho_c &=& \prod_{\otimes_s} \Bigl[(1- \lambda_s)~c^\dagger_s \ket{0}\!\bra{0} c_s +  \lambda_s ~ \ket{0}\! \bra{0}\, \Bigr] \equiv \prod_{\otimes_s} (\rho_c)_s
\label{vN12}
\eeqar
There is a slight abuse of notation here in continuing to use $\ket{0}$. It should be
noted that, in $\rho_b$, while the state $\ket{0}\!\bra{0}$, which is obtained by tracing over the
$c$'s, has no $b$-occupancy, it is not empty. It stands for
$\sum_s c_s^\dagger \ket{0, \tilde{0}} \bra{0, \tilde {0}} c_s$ if we 
consider a more elaborate notation of two copies of $\H$ as in section 2.
Thus $\ket{0}\! \bra{0}$ in the first line of (\ref{vN12})
does capture the effect of fermions outside $\vert z\vert =1$, although the effect is small, since
$1- \lambda_s$ will be small for states localized far into the outside region, i.e., for
states with $s \gg {\half} n$. And a similar statement, {\it mutatis mutandis}, holds for
the second line of (\ref{vN12}) as well.
In each of the cases in (\ref{vN12}) one can define the von Neumann entropy, which is also the entanglement entropy, 
\beqar
S&=& -\Tr ( \rho_b \, \log \rho_b ) = -\Tr ( \rho_c \, \log \rho_c )
\nonumber\\
&=& 
- \sum_s \left[ \lambda_s \log \lambda_s + (1-\lambda_s) \log (1- \lambda_s)\right]
\label{vN13}
\eeqar
This method of splitting $a_s, \,a^\dagger_s$ as in (\ref{vN7}) and calculating the entanglement
entropy was first given in \cite{dubail}. For entanglement entropy for
the $\nu =1 $ state in two dimensions, see also \cite{rodr}.
Some of the other references on the entanglement entropy for Hall systems
are given in \cite{{other1}, {other2}}.

We now want to consider the modular operator corresponding to the states $\rho_b$, $\rho_c$.
A general state can be taken to be of the form
\beq
\Phi = \prod_{\otimes_s} \left[ \begin{matrix}
\alpha +\beta & \gamma -i \delta\\
\gamma+i \delta & \alpha -\beta\\ \end{matrix} \right]_s
\label{vN14}
\eeq
with arbitrary elements $\alpha, \cdots, \delta$. In this notation, the $(11)$ element
corresponds to $b^\dagger\ket{0} \bra{0} b$, $(12)$ to
$b^\dagger \ket{0} \bra{0} c$, $(21)$ to $c^\dagger \ket{0} \bra{0} b$, 
$(22)$ to $c^\dagger \ket{0} \bra{0} c$. Following \cite{witten} and our discussion in section 2, we can define the
action of the modular operator $\Delta$ as
\beq
\Delta \,  \Phi = \rho_b \, \Phi \, (\rho_c)^{-1}
\label{vN15}
\eeq
Just to reiterate, in this expression,
the state dependence of $\Delta$ is given in terms of
$\rho_b$, $\rho_c$.
For each $2\times 2$ subspace, (\ref{vN15}) works out to
\beq
\Delta \, \left( \begin{matrix} \alpha\\ \beta\\ \gamma\\ \delta\\ \end{matrix}\right)
= \left[ \begin{matrix} 
1&0&0&0\\ 
0&1&0&0\\ 
0&0& {1\over 2} (x + x^{-1}) & -{i\over 2} (x- x^{-1})\\
0&0& {i\over 2} (x - x^{-1}) & {1\over 2} (x+ x^{-1})\\
\end{matrix}\right] \left( \begin{matrix} \alpha\\ \beta\\ \gamma\\ \delta\\ \end{matrix}\right)
\label{vN16}
\eeq
where $x = \lambda/(1-\lambda)$. 
The eigenvalues are $1$, $1$, $\lambda/(1-\lambda)$, $(1-\lambda)/\lambda$.
This is for one value of $s$. The full spectrum is thus given by
the product of these eigenvalues over all values of $s$.
Thus
\beq
{\rm Spec}(\Delta ) = \left( \{1\}, \{1\}, \{ \lambda_s /(1- \lambda_s)\} , \{ {(1-\lambda_s ) /\lambda_s}\},
\{  \lambda_{s_1}  \lambda_{s_2} /(1- \lambda_{s_1} ) /(1- \lambda_{s_2}) \}, {\rm etc.}
\right)
\label{vN17}
\eeq
The products of the individual eigenvalues for all
values of $s$ are included in this set.
 Since some of the eigenvalues are just $1$, the individual eigenvalues
get repeated as well in this set.

Our first result is to show that, as $n \rightarrow \infty$ for states on $S^2$, for
any value between
zero and $1$, there is
some $k$ such that $\lambda_s$ is equal to this chosen value.
This will imply that the spectrum of $\Delta$ is
the interval $[0, \infty )$.
The calculations are given in Appendix A. We show that the values of
$\lambda_s$ start near $1$ for $s\ll n$ and drops to zero as
$s$ becomes close to $n$. The maximal difference of $\lambda_s$ for nearby values
of $s$ occurs at $s = \half n$, where 
\beq
{\lambda_{{n\over 2}}  - \lambda_{{n\over 2}+1} 
\over \lambda_{{n\over 2}} } =\sqrt{8\over \pi}\, {1\over \sqrt{n}} + {\cal O}(1/n)
\label{vN21}
\eeq
We can therefore conclude that the differences between $\lambda_s$ and $\lambda_{s+1}$ vanish
for all $s$, as $n \rightarrow \infty$, showing that the values of $\lambda_s$
 fill the interval
between zero and $1$.
In other words, the spectrum of
$\Delta$ for this state is ${\mathbb R}_+$.

For the analysis given above, we chose the dividing line between the two regions as the equator, at $\vert z\vert = 1$.
The result can be generalized to an arbitrary value of $\vert z\vert$ for the dividing line, so long as the number of states
in each region tends to infinity as $n \rightarrow \infty$. Consider 
$\vert z\vert = R$. The relevant integral is now
\beq
\lambda_s = {(n+1)! \over s!\, (n-s)!}
\, \int_0^{R^2} du {u^s \over (1+u)^{n+2}}
\label{vN22}
\eeq
The middle of the transition region between $\lambda = 0$ and $\lambda = 1$
will occur at $s = s_*= R^2 n /(1 +R^2)$. 
In this case, the maximal difference is given by
\beq
{\lambda_{s_*} - \lambda_{s_* +1} \over \lambda_{s_*} } 
\approx \sqrt{ 2 \over w (1-w ) \pi} ~{1 \over \sqrt{n} } + \cdots, \hskip .3in 
w = {R^2 \over 1+ R^2}
\label{vN23}
\eeq
Once again, we notice that 
the differences of the nearby $\lambda$'s vanish as $n \rightarrow \infty$. Thus
the values of $\lambda$ will fill the interval between zero and $1$, leading to the spectrum of
$\Delta$ as ${\mathbb R}_+$.

Deformations of the dividing line can be viewed as area preserving diffeomorphims which are
realized in terms of unitary transformations of the lowest Landau levels states.
The spectrum of $\Delta$ will not be sensitive to this, so the conclusion holds more generally than
for the case of a circular dividing line.

If we consider the lowest Landau level as a model for the fuzzy version of
$S^2$, then the relevant state must be the fully filled level with $\nu = 1$.
Having unfilled one-particle states will correspond to having
the two-sphere with points removed, as $n \rightarrow \infty$.
Therefore the only set of states relevant for the case of fuzzy $S^2$
will correspond to different choices of $R$. This leads to the conclusion:
\begin{quotation}
\noindent For all allowable states in the framework of using the lowest Landau as a model for a 
noncommutative space,
$\bigcap_{\Psi } {\rm Spec} (\Delta_\Psi ) = {\mathbb R}_+$.
\end{quotation}

\section{Generalization to ${\mathbb {CP}}^k$}

The results we have obtained for $S^2 \sim {\mathbb{CP}^k}$ can be easily generalized
to ${\mathbb{CP}^k}$, $k > 1$ \cite{KN1}. We may view this space as a group coset,
\beq
{\mathbb{CP}^k} = {SU(k+1) \over U(k)}
\label{vN24}
\eeq
This is a homogeneous space with $U(k)$ as the isotropy group.
The curvature is thus valued in the Lie algebra of $U(k)$ and is constant in 
the tangent frame basis. This means that we can introduce additional 
gauge fields with the field strength proportional to the curvatures and thus set up
the analogue of the Landau problem and Hall effect.
More explicitly,
the wave functions 
can be considered
as functions on 
$SU(k+1)$ which have a specific transformation property under
the $U(k) \subset SU(k+1)$.
A basis for functions on the group $SU(k+1)$
is given by  the
matrices corresponding to the
group elements in the unitary irreducible representations, or the so-called
Wigner $\cal{D}$-functions, which are defined as
\beq
\D^{(J)}_{\mathfrak{l}; \mathfrak{r}}(g) = \la J ,\mathfrak{l}\vert\, g\,\vert J, \mathfrak{r} \ra \label {5}
\eeq
where $J$ denotes the irreducible representation
and $\mathfrak{l}, ~\mathfrak{r} $ stand for two sets of quantum numbers specifying the 
states within the representation.
There is a natural left and right
action of group translations on an element $g\in SU(k+1)$, 
defined by
\beq
{\hat{L}}_A ~g = T_A ~g, \hskip 1in {\hat{R}}_A~ g = g~T_A
\label{6}
\eeq
where $T_A$ are the $SU(k+1)$ generators in the representation to which $g$ belongs.

The generators of $SU(k+1)$ which are not in
the algebra of $U(k) \subset SU(k+1)$ can be separated into
$T_{+i}$, $i=1,2 \cdots ,k$, which are
of the raising type and $T_{-i}$ which are of the lowering
type. 
These generate translations while $U(k)$ generates rotations at a point.
The covariant derivatives on ${\mathbb{CP}}^k$
are given by
\beq
\D_{\pm i}  = i\,{{\hat R}_{\pm i} \over r}
\label{6a}
\eeq
where $r$ is a parameter with the dimensions of length.
(The volume of the manifold will be proportional to $r^{2k}$.)
The strength of the gauge field should be given  by the commutator of covariant
derivatives. The commutators of 
${\hat R}_{+i}$ and ${\hat R}_{-i}$ are in the
Lie algebra of $U(k)$,
so we can specify the background field by
specifying the right action of $U(k)$ on the wave functions.
For the constant background field, the relevant conditions are
\beqar
{\hat R}_a ~\Psi^J_{m; \alpha} (g) &=&
(T_a)_{\alpha \beta} \Psi^J_{m; \beta} (g) \label{9a}\\
{\hat R}_{k^2 +2k} ~\Psi^J_{m; \alpha} (g) &=& - {n k\over \sqrt{2 k
(k+1)}}~\Psi^J_{m; \alpha} (g) \label{9b}
\eeqar
where $m$ ($=1,\cdots, {\rm dim}J$) counts the degeneracy of the Landau level.
Equation (\ref{9a}) shows that the wave functions $\Psi^J_{m; \alpha}$ transform,
under right
rotations, as a representation ${\tilde J}$
of $SU(k)$. Here
$(T_a)_{\alpha \beta}$ are the representation matrices for the
generators of
$SU(k)$ in the representation ${\tilde J}$, and
$n$ is an integer characterizing the Abelian part of the background field.
$\alpha ,\beta$ label states within the $SU(k)$ representation ${\tilde J}$
(which is itself
contained in the representation $J$ of $SU(k+1)$). The index $\alpha$ carried by the
wave functions $\Psi^J_{m; \alpha} (g)$
 is basically the gauge index. The wave functions are sections
of a $U(k)$-bundle on ${\mathbb{CP}}^k$. In terms of $\D$-functions, they are given by
$\Psi^J_{m; \alpha} (g) = \sqrt{{\rm dim}J} \, \bra{J, m} g \ket{J, \alpha, n}$.

The Hamiltonian $H$ for the Landau problem is proportional to the covariant Laplacian on 
${\mathbb{CP}}^k$; explicitly the action of $H$ on wave functions is given by
\beq
H \, \Psi = - {1\over 4 m} (\D_{+i} \D_{-i} + \D_{-i} \D_{+i} ) \, \Psi
\label{9c}
\eeq
Since the commutator of $ [{\hat R}_{+i}, {\hat R}_{-i}]$ is in the algebra of 
$U(k)$, we see from (\ref{6a}) and (\ref{9b}) that
$H$ is proportional to 
$\sum_{i} {\hat R}_{+i} {\hat R}_{-i}$, apart from additive constants.
Thus the lowest Landau level should satisfy, in addition to the
requirements (\ref{9a}, \ref{9b}), the condition
\beq
\hat{R}_{-i} \, \Psi = 0
\label{9d}
\eeq
This is the holomorphicity condition on the lowest Landau level wave functions.

We consider, for simplicity, the case of a $U(1)$ background, taking $\ket{J, \mathfrak{r}}$
to correspond to the trivial (singlet)  representation for $SU(k) \in SU(k+1)$.
The relevant representations are then the rank $n$ totally symmetric representations
of $SU(k+1)$ and we can construct them explicitly using complex coordinates
for ${\mathbb{CP}^k}$ as
\beq
\Psi_{i_1 i_2 \cdots i_k} = \sqrt{N}\,
\left[ {n! \over i_1! i_2! \cdots i_k! (n-s)!}\right]^{\half}\,
{z_1^{i_1} z_2^{i_2}\cdots z_k^{i_k} \over (1 + \bz\cdot z)^{n \over 2}}
\label{vN39}
\eeq
where $s = i_1 + i_2 +\cdots+ i_k$ and $N = {\rm dim}J = (n+k)!/ (n! k! )$ is the 
total number of states or degeneracy
of the LLL.
The volume element for ${\mathbb{CP}^k}$ is 
\beq
d\mu = {k! \over \pi^k} {d^2z_1 \cdots d^2z_k
\over (1+ \bz \cdot z)^{k+1}}
\label{vN40}
\eeq
We have chosen the normalization such that the total volume, $\int d\mu$, is $1$.
For the entanglement entropy, we thus need
\beqar
\lambda_{i_1 i_2 \cdots i_k}  &=& \int_0^R d\mu \,\Psi^*_{i_1 i_2 \cdots i_k}  \Psi_{i_1 i_2 \cdots i_k} 
\nonumber\\
&=& {(n+k)! \over (s+k -1)! (n-s)!}\,
\int_0^{R^2} du {u^{s+k -1} \over (1+u)^{n+k+1}}
\label{vN41}
\eeqar
where, in the second line, we have carried out the angular integrations taking the
interface to be spherically symmetric. The maximal difference of $\lambda_s$ nearby
values of $s$ is now obtained for $s= s_* = w (n+k -1) - (k-1)$, and
\beq
{\lambda_{s_*} - \lambda_{s_* +1} \over \lambda_{s_*}}
= \sqrt{2 \over w (1-w) \pi}  \, {1\over \sqrt{n+k-1}} + \cdots, \hskip .2in
w = {R^2 \over 1+R^2}
\label{vN42}
\eeq
As before, we then find that all values between zero and $1$ are realized for
some $\lambda_s$, leading to the same conclusion:
\begin{quotation}
\noindent For all allowable states in the framework of using the lowest Landau as a model for a 
noncommutative version of ${\mathbb{CP}^k}$,
$\bigcap_{\Psi } {\rm Spec} (\Delta_\Psi ) = {\mathbb R}_+$.
\end{quotation}
We have shown this result only for the case of a $U(1)$ background. We expect this to hold even with additional nonabelian background fields. This will be taken up in a subsequent paper.

\section{Arbitrary background fields and spin connection}

We now turn to the second issue mentioned in the introduction, namely, the dependence of
the entanglement entropy on the background fields and the spin connection. For this, we need to
know how $u_s^* u_s$ depends on these quantities. One way to identify this
dependence is to write the
Hamiltonian (\ref{9c}) where the covariant derivatives include additional background fields 
and then use
perturbation theory to calculate the change in $u_s^* u_s$. A limit of
$m \rightarrow 0$ may then be taken at the end to isolate the LLL wave functions.
A simpler alternative, which we shall consider here, is to utilize
previous calculations for the effective action for droplets of fermions \cite{KN1}.
The basic strategy is the following. We will consider a general occupancy matrix
for a subset of the states of the lowest Landau level. We can then define a function
similar to the symbol for this matrix which carries information about the wave functions.
A general ansatz for this function can then be written down.
Using an index theorem appropriate to the states of the LLL and considering special cases
we can firm up the various terms in the ansatz. This will then yield the leading terms for
the background dependence
of $u_s^* u_s$. This part of the reasoning will rely on
\cite{KN-Dolb} where the
Dolbeault index density was used to obtain the bulk effective action for
the $\nu = 1$ state. 

Before proceeding to the main line of reasoning, we assemble two key ingredients, 
namely, the index theorem and the generalized Chern-Simons form.
The wave functions of the lowest Landau level obey a holomorphicity condition,
which is (\ref{9d}) for $\mathbb{CP}^k$ and a suitable generalization of
the same for other complex manifolds. The background fields are included
in the relevant antiholomorphic derivatives via
conditions like (\ref{9a}), (\ref{9b}). Thus we are looking for the kernel of the
antiholomorphic covariant derivatives on $\M$. This is given by the Dolbeault index,
with the
index density
\beqar
\I_{\rm Dolb} &=& {\rm td}(T_c \M) \wedge {\rm ch} (V)\nonumber\\
&=&{1\over 2 \pi} \Tr \left( F + {\half} {R }\right)
- {1\over 8 \pi^2}\left[
\Tr F^2 + \Tr R \, \Tr F + {1\over 4} (\Tr R )^2 - {1\over 12} \Tr R^2 \right] +\cdots
\label{vN55}
\eeqar
where $F$ is the two-form field strength for the gauge field and $R$ is the curvature two-form.
In this equation ${\rm td} (T_c \M)$ is the Todd class on the 
complex tangent space of the manifold
$\M$. Rather than give the general formula using a splitting principle,
we display the expansion in powers of the curvature. 
The general formula is given in
\cite{EGH} and is discussed in the specific context of
Landau level states in
\cite{KN-Dolb}.
Also, in (\ref{vN55}) ${\rm ch} (V) = \Tr (e^{iF/2\pi} )$ is the Chern character of the vector bundle $V$.
(The charged fields defining the Landau problem
are sections of this vector bundle; i.e., they have ${\rm dim}V$ components corresponding to the
representation for the nonabelian gauge group, each component being
a local function on $\M$. For example, for ${\mathbb{CP}^k}$, we can consider
fields $\phi_\alpha$ with wave functions of the form
$\Psi^J_{m; \alpha} (g) = \sqrt{{\rm dim}J} \, \bra{J, m} g \ket{J, \alpha, n}$ as in section 4.
In this case, ${\rm dim}V = {\rm dim}{\tilde J}$ which is the dimension
of the $U(k)$ representation carried by
the state $\ket{J, \alpha, n}$. For more details, see \cite{KN-Dolb}.)
As with such formulae, for two-dimensional manifolds we use the two-form part of
$\I_{\rm Dolb} $ from (\ref{vN55}), for four-manifolds we use the four-form part, etc.

Turning to the second ingredient, namely, the generalized Chern-Simons form,
notice that the index densities involve symmetrized traces of powers of
the gauge field strength (as a two-form), or powers of the curvature two-form,
or mixed terms involving products of powers of both.
Now, if
$\P(F)$ is an invariant polynomial which is the symmetrized trace of a product of $k$ 
$F$'s, then
\beqar
\P (F^{(1)}) - \P(F^{(2)}) &=& d \, Q(A^{(1)}, A^{(2)})\nonumber\\
Q(A^{(1)}, A^{(2)})&=& k \int_0^1 dt~ \P (A^{(1)}-A^{(2)}, F_t, F_t, \cdots, F_t),
\label{vN45}
\eeqar
where $A_t = A^{(2)} + t (A^{(1)} - A^{(2)})$, with $F_t =
dA_t + A_t^2$.
 Notice that $A_t$ defines a straight line
in the space of potentials connecting the two potentials
$A^{(2)}$ and $A^{(1)}$. 
Equation (\ref{vN45}) is the definition of the generalized Chern-Simons term
$Q(A^{(1)}, A^{(2)})$ \cite{EGH}. This version of the Chern-Simons form
has been used in physics contexts before, for example, in obtaining
expressions for gauge (and gravitational) anomalies
with a nontrivial gauge (or gravitational) background \cite{MSZ}. We may also note that, since
$\P(F)$ can be written as the derivative of a Chern-Simons term
${\rm C.S.}(A)$, (\ref{vN45}) is equivalent to writing
\beq
Q(A^{(1)}, A^{(2)}) = {\rm C.S.}(A^{(1)}) - {\rm C.S.} (A^{(2)}) + d  B(A^{(1)}, A^{(2)})
\label{vN46}
\eeq
The formula (\ref{vN45}) has the advantage of
providing an explicit expression for
$B(A^{(1)}, A^{(2)})$ as well. Even though we used the notation of $A$ and $F$ in
equations (\ref{vN45}, \ref{vN46}),
the statements equally well apply to the spin connection and the curvature.

We now consider the case of $M$ of the lowest Landau levels being occupied.
(Eventually, we will be interested in the $\nu = 1$ state with all states being occupied.
What we outline here with a smaller droplet is only a trick to get the background dependence of
$u_s^* u_s$.)
We can specify the droplet of $M$ occupied  states by the occupancy matrix
$P$ which is given by
\beq
P_{ij} = \begin{cases} \delta_{ij} , \hskip .2in &i, j = 0, 1,\cdots, (M-1)\\
0 \hskip .2in &i, j \geq M
\end{cases}
\label{vN44}
\eeq
Corresponding to this occupancy matrix, we introduce the function\footnote{This is related to 
what is called the symbol for $P$ by a factor of $N$, the total degeneracy of the LLL.  The symbol
is defined using just the group elements, say $ \bra{J, m} g \ket{J, \alpha, n}$ for $\mathbb{CP}^k$.
Thus the symbol of
$P$ is ${1\over N} (P)$ \cite{KN1}, \cite{KNrev}.}
\beq
(P)_{M-1} =  \sum_0^{M-1} u_s^* u_s
\label{vN44a}
\eeq
Thus we may write
$u_M^* u_M =  (P)_M - (P)_{M-1}  $. The function $(P)$ is basically the number density of
particles in the lowest Landau level, and hence it should be proportional to the index density for the
$\nu =1$ state, modulo terms which integrate to zero. This is the point of utility of
the index theorem.

The background field dependence of $u_s^* u_s$ involves the comparison of two choices of the background. We will denote the background fields we start with by
$a$ and $\omega^{(0)}$, where $a$ is the potential for the gauge part and
$\omega^{(0)}$ is the spin connection.
For example, for ${\mathbb{CP}}^k$, the Abelian part of 
$a$ will be an appropriate multiple of the K\"ahler form; there can be a nonabelian
background as well. The spin connection
$\omega^{(0)}$ will correspond to the standard curvatures for
${\mathbb{CP}}^k$ with the Fubini-Study metric.
The general background we want to consider will have gauge fields
$a + A$ and spin connection $\omega$.
We are thus interested in the function $(P)$ calculated with
the one-particle wave functions corresponding to the background
$(a, \omega^{(0)})$  and with those corresponding to the background
$(a +A, \omega )$.
We can then use the result
$u_M^* u_M =  (P)_M - (P)_{M-1} $ to identify the background dependence.

An important point is that
the new values of the background fields, i.e.,  $(a +A, \omega )$,
must be such that the total number of states
obtained by quantization
remains the same. 
Thus the fields we are considering must all be in the same topological class, so
that the Dolbeault index is unaltered by $a \rightarrow a+A, \, 
\omega^{(0)} \rightarrow \omega$. In this sense, we may think of
the new fields as a perturbation of the old ones.

We can now write down a general ansatz for $(P)$ as
\beq
{\widetilde{(P)} } = {\I_{\rm Dolb} \over N} ~ (P)^{(0)} - K \, d (P)^{(0)} + d {\mathbb X}
\label{vN47}
\eeq
Here $(P)^{(0)}$ denotes the function corresponding to
$P$ calculated with the unperturbed
one-particle wave functions, namely, with the background
$A^{(2)} = (a, \omega^{(0)})$. On the left hand side, we have the function for $P$
calculated with the perturbed wave functions corresponding to
$A^{(1)} = (a+A, \omega )$. We actually use the dual on the left hand side
so that it can be viewed as a $2k$-form;
this is signified by the
tilde sign. ${\widetilde{(P)} }$ is to be viewed as the function
$(P)$ multiplied by the volume form appropriate to the 
background $(a+A, \omega )$. Further,
$K$ is a $(2k -1)$-form, so is ${\mathbb{X}}$. 
The nature of the terms in (\ref{vN47}) and the justification for them can be seen from
the following observations.
\begin{enumerate}
\item First consider the case where all states are filled, so that $M = N$.
When all states are filled, $(P)^{(0)}$ is a constant,
in fact equal to $N$ for large $n$,
as seen from \cite{KN1}, \cite{ray-sakita}.
(The symbol is equal to $1$,
but since we have used the normalized wave functions, $(P)^{(0)} = N$ in this case.)
We see that (\ref{vN47}) implies that ${\widetilde{(P)} } = \I_{\rm Dolb}$,
provided $d\mathbb{X}$ also involves only derivatives of
$(P)^{(0)}$.
The result ${\widetilde{(P)} } = \I_{\rm Dolb}$ is as it should be, 
since  ${\widetilde{(P)} }$ is the number density of the
occupied states and it should be the index density when all states are occupied.
\item Secondly, consider the case when the additional background fields are zero,
i.e., we have only $(a, \omega^{(0)})$, but keeping $M < N$. In this case, we expect
${\widetilde{(P)} } = d\mu\, (P)^{(0)}$. In (\ref{vN47}), we can thus set
$\I_{\rm Dolb} = \I_{\rm Dolb} (a, \omega^{(0)})$. Further, we have
\beq
\I_{\rm Dolb} (a, \omega^{(0)}) = N\, d\mu
\label{vN48}
\eeq
The factor of $N$ in this formula is easily understood.
It is needed to ensure that the integral of
$\int \I_{\rm Dolb} (a, \omega^{(0)}) $ gives $N$ for the unperturbed case,
since we normalized the
unperturbed volume element to integrate to $1$.
We see that (\ref{vN47}) consistently reduces to
${\widetilde{(P)} } = d\mu\, (P)^{(0)}$, provided
both $K$ and $\mathbb{X}$ vanish when the additional gauge fields
are set to zero.
\item Continuing with the case of $M < N$, we expand $\I_{\rm Dolb}$
around $(a, \omega^{(0)})$ using (\ref{vN45}), i.e.,
\beq
\I_{\rm Dolb}  (a+A, \omega) = \I_{\rm Dolb} (a, \omega^{(0)} ) + d Q 
\label{vN49}
\eeq
Using this relation, (\ref{vN47}) becomes
\beq
{\widetilde{(P)} }= d\mu ~ (P)^{(0)} + {d Q\over N} \, (P)^{(0)} - K \, d (P)^{(0)} + d {\mathbb X}
\label{vN50}
\eeq
Since the total number of states should be the same for both
$(a, \omega^{(0)} )$ and $(a+A, \omega)$, terms
in (\ref{vN47}) other than $d\mu ~ (P)^{(0)}$
must combine into a total derivative, so that they can give zero
upon integration. This identifies
$K = Q/N$.
\item Finally, we have already mentioned that ${\mathbb{X}}$
must vanish when the additional gauge fields are set to zero
and that it should also vanish when $(P)^{(0)}$ is a constant, from items
2 and 1 above.
Therefore $\mathbb{X}$ can be written as
$W^i \del_i (P)^{(0)}$, where $W^i$ is a $(2k-1)$-form and is also
a vector. It is not determined by our arguments so far. 
\end{enumerate}
Collecting all these
results together, we can now rewrite (\ref{vN47}) as
\beq
{\widetilde{(P)} } = d\mu ~ (P)^{(0)}  + d \left[ {Q~ P^{(0)} \over N}
\right] +  d \left( W^i \del_i(P)^{(0)} \right)
\label{vN51}
\eeq
where $W^i $ is zero when restricted to $a, \omega^{(0)}$.
As mentioned before, the relevant index density we should use for
the generalized Chern-Simons form $Q$ will be
the Dolbeault index density
$\I_{\rm Dolb}$, with
$A^{(1)} = (a +A, \omega )$ and $A^{(2)}  = (a, \omega^{(0)})$.

Taking the difference between $(P)_M$ and $(P)_{M-1}$, we conclude
that
\beq
d\mu~ \bigl[u^*_su_s\bigr]_{a+A, \omega} = d\mu~ u^*_su_s
+ d \left[ {Q ~u^*_su_s\over N}
\right] +  d \left( W^i \del_i(u^*_su_s) \right)
\label{vN52}
\eeq
$u^*_s u_s$ in all terms on the right hand side is evaluated with the
unperturbed background $(a, \omega^{(0)})$.
Going back to the fully filled states, i.e., $\nu = 1$, we can now integrate
(\ref{vN52}) over a region  $D$ whose boundary corresponds to the entangling surface.
The result is
\beq
\bigl[ \lambda_s \bigr]_{a+A, \omega} = 
\bigl[\lambda_s\bigr]_{a, \omega^{(0)}} +
{1\over N} \oint_{\del D} Q \, \bigl[u^*_s u_s\bigr]_{a, \omega^{(0)}} 
+ \oint_{\del D} W^i \bigl[\del_i(u^*_s u_s) \bigr]_{a, \omega^{(0)}} 
\label{vN53}
\eeq
The $\lambda_s$-terms in this equation is the result after integrating over the boundary
$\del D$; i.e., the analogue of the angular integrations has been carried out as
discussed earlier and in the appendixes.

We can now use (\ref{vN53}) for the entropy.
In the spirit of perturbation theory, the change in the entanglement entropy due to the
change in the background fields is given by
\beqar
S&=& S(a, \omega^{(0)}) - \sum_s \log\left({\lambda_s \over 1- \lambda_s}\right)
\delta \lambda_s\nonumber\\
&=& S(a, \omega^{(0)}) - \oint_{\del D} {Q\over N} 
\sum_s \log\left({\lambda_s \over 1- \lambda_s}\right) \,
\bigl[u^*_s u_s\bigr]_{\del D; a, \omega^{(0)}} \nonumber\\
&&\hskip .2in - \sum_s \log\left({\lambda_s \over 1- \lambda_s}\right) 
\oint_{\del D} W^i \bigl[\del_i(u^*_s u_s) \bigr]_{\del D; a, \omega^{(0)}} 
\label{vN54}
\eeqar
This is the main result of this section,
summarizing our expectation for the background dependence
in terms of $Q$. The precise form of the remaining factors is not important
regarding the background dependence.
The arguments which led to this result are
very general, but indirect, based on index theorems.
We will carry out an explicit calculation, which is presented in
Appendix B, for some special cases. (We expect to present the explicit calculations for
the more general cases in a separate paper.)
This will show that
the second correction from (\ref{vN53}), namely,
$W^i \bigl[\del_i(u^*_s u_s) \bigr]_{a, \omega^{(0)}} $ may be taken to
be subdominant compared to the first, in some qualified sense.
Taking this into account, we may restate the result as follows.
\begin{quotation}\noindent
The leading term in the
dependence of the entanglement entropy on the gauge fields and spin connection
is proportional to
the generalized Chern-Simons term $Q$ for the Dolbeault index density.
\end{quotation}

It may be useful at this stage to see the explicit formulae for the
generalized Chern-Simons forms
relevant to some lower dimensional examples, rather than the more cryptic 
expression (\ref{vN45}). Using (\ref{vN55}) and (\ref{vN45}) we find
\beqar
Q_{2d}&=&{1\over 2 \pi} \Tr\left( A + {\half} (\omega - \omega^{(0)}) \, \right)\label{vN56}\\
Q_{4d}&=& - {1\over 8 \pi^2} \Bigl[
-4\pi\, {\rm C.S.} (a +A ) +4\pi \, {\rm C.S.} (a)  + d\, \Tr (a A) \nonumber\\
&&\hskip .5in - {1\over 12} \left( -4\pi \,{\rm C.S.} (\omega) +4\pi \, {\rm C.S.} (\omega^{(0)})  + d\, \Tr \bigl(\omega^{(0)} (\omega -
\omega^{(0)}) \bigr) \right) \nonumber\\
&&\hskip .5in + \Tr (\omega - \omega^{(0)} )\,\Tr(dA) 
+ \Tr (\omega - \omega^{(0)}) \,\Tr(da + a^2) + \Tr (A) \, \Tr (d\omega^{(0)})\nonumber\\
&&\hskip .5in + {1\over 2} \Tr (\omega - \omega^{(0)})\, \Tr (d\omega^{(0)})
+ {1\over 4} \Tr (\omega - \omega^{(0)})\,\Tr(d \omega - d\omega^{(0)} )
\Bigr]
\label{vN57}
\eeqar
Here ${\rm C.S.}$ stands for the Chern-Simons three-form, given,
for a generic argument $A$, as
\beq
{\rm C.S.} (A) = - {1\over 4\pi} \Tr \left( A dA + {2\over 3} A^3 \right)
\label{vN58}
\eeq

\section{Discussion}

The main results of this paper are about the spectrum of the modular operator
and the dependence of the entanglement entropy on the background fields
and spin connection, for a noncommutative space, or equivalently, for
the $\nu = 1$ quantum Hall state. These have been spelled out
at the end of sections 3 and 5.
An important direction to explore further is the
term involving $ \left( W^i \del_i(u^*_su_s) \right)$ in (\ref{vN51}), (\ref{vN52}).
This is presumably related to the boundary actions for a droplet of finite size.
In the context of quantum Hall effect, while any direct experimental implication
is unclear, our result is in the nature of elucidating
general properties of Hall states, 

As for the context of noncommutative geometry, the following comments
may help with the placement of our
results in a larger context.
In the introduction we have already alluded to the nexus of ideas about entropy and gravity,
entanglement as an integral feature of relativistic field theory, and noncommutative geometry
which attributes degrees of freedom to space itself.
To this we may add the observation that, in 2+1 dimensions,
standard Einstein gravity can be described by an action which is
the difference of two Chern-Simons terms \cite{3dgrav}.
In higher dimensions, one can consider a class of gravity theories
with Chern-Simons actions, although they do not correspond to the standard
Einstein gravity \cite{zanelli}. In all these cases, the field equations of gravity arise
as extremization of an action which is a combination of Chern-Simons terms.
Our result
that the leading term in the background dependence of 
the entanglement entropy in noncommutative geometry (modeled as quantum Hall systems)
is given by a generalized Chern-Simons term takes on added significance when viewed
within this circle of ideas.
We expect that this result can be utilized to develop an approach to gravity in odd dimensional spacetimes
based on noncommutative geometry by modeling space by quantum Hall systems
and that such a description would naturally realize the field equations
for gravity as maximization conditions for the entanglement entropy.
We plan to explore this idea further in future publications.

\section*{Appendix A: Asymptotic formulae for $\lambda_s$}
\def\theequation{A\arabic{equation}}
\setcounter{equation}{0}

We will start with equation (\ref{vN41})  from text, which gives
\beq
\lambda_s =  {(n+k)! \over (s+k -1)! (n-s)!}\,
\int_0^{R^2} du {u^{s+k -1} \over (1+u)^{n+k+1}}
\label{B1}
\eeq
This formula applies to $\mathbb{CP}^k$. We started in the text with the case
of $k = 1$ for $\mathbb{CP}^1 \sim S^2$ and with $R= 1$. Those results can be obtained as
special cases of the formulae give here.

We first consider small values of $s$ compared to $n$. 
The integrand in (\ref{B1}) is a function which peaks around some value of
$u$. This value is near zero for small $s$,
moving to large values of $u$ as $s$ becomes large.
 Thus, for small values of $s$ compared to $n$, $\lambda_s$ will be close to $1$.
 The full integral, up to infinity, gives $1$, so we can rewrite (\ref{B1}) as
\beq
\lambda_s =  1- \,{(n+k)! \over (s+k -1)! (n-s)!}\,
\int_{R^2}^\infty du {u^{s+k -1} \over (1+u)^{n+k+1}}
\label{B2}
\eeq
Making a change of variables $u = x/(n+k+1)$, we find
\beqar
\lambda_s &=& 1 - \,{(n+k)! \over \Gamma(s+k) (n-s)!}\, {1\over (n+k+1)^{s+k}}
\nonumber\\
&&\hskip 1in \times \int_{R^2(n+k+1)}^\infty dx \,x^{s+k -1} e^{-x}\left( 1 + {x^2 \over 2 (n+k+1)} + \cdots\right)\nonumber\\
&\approx& 1 - {\Gamma(s+k, R^2 (n+k+1)) \over \Gamma(s+k)} + \cdots
\nonumber\\
&\approx& 1 - {\left[ R^2 (n+k+1)\right]^{s+k-1} \over \Gamma(s+k)} \, e^{- R^2 (n+k+1)} + \cdots
\label{B3}
\eeqar
where, in the first line, we used 
\beq
\left( 1 + {x \over N} \right)^N \approx e^x \left( 1 - { x^2 \over 2 N} +\cdots
\right), \hskip .2in {\rm as}~ N \rightarrow \infty .
\label{B4}
\eeq
For the second and third lines we used the definition of the incomplete $\Gamma$-function
and its asymptotic expansion \cite{AS},
\beqar
\Gamma(s+k, X ) &=& \int_X^\infty dx\, x^{s+k -1} e^{-x} 
\nonumber\\
&\approx& X^{s+k -1} e^{- X} ~+~ \cdots
\label{B5}
\eeqar
Equation (\ref{B3}) shows that $\lambda_s$ is exponentially close to $1$ for
small
values of $s$, as $n$ becomes large. The ellipsis indicates terms which are smaller than 
what is displayed.

For values of $s$ close to $n$, we can do a similar analysis. Writing
$s = n-r$ and carrying out an inversion $u = (n+k+1)/x$, we find
\beqar
\lambda_{n-r}&=& {(n+k)!\over \Gamma(n-r+k)\,  r!}
{1\over (n+k+1)^{r+1}} \nonumber\\
&&\hskip 1in \times\int_{(n+k+1)\over R^2}^\infty dx\, x^r e^{-x} \left( 1
+ {w^2 \over 2 (n+k+1)} +\cdots\right)\nonumber\\
&\approx& {1\over \Gamma(r+1)} \left({n+k+1 \over R^2}\right)^r \, e^{- (n+k+1)/R^2} + \cdots
\label{B6}
\eeqar
We see that, for $s$ near $n$, the values are exponentially small.

\begin{figure}[b]
\begin{center}
\begin{minipage}{5cm}
\scalebox{.4}{\includegraphics{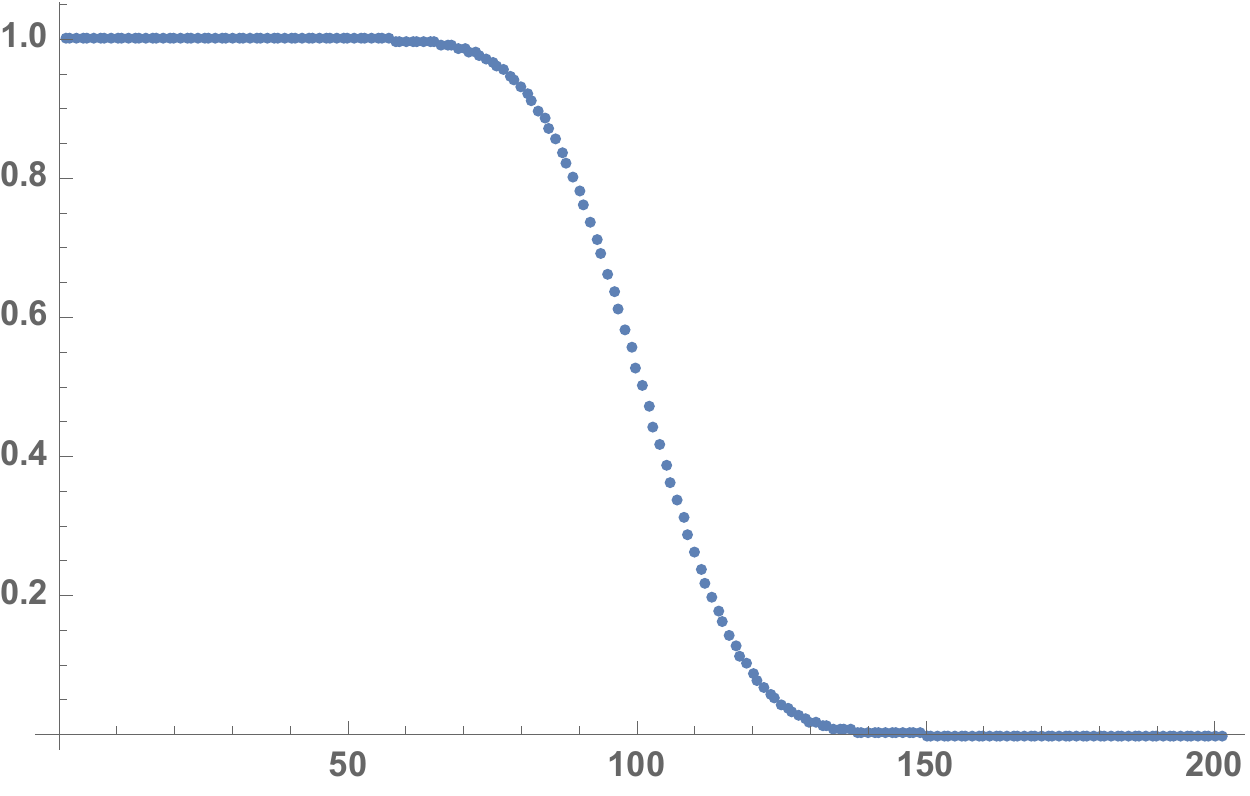}}
\caption{Values of $\lambda_s$ as a function of $s$, for $n = 200$}
\label{g1}
\end{minipage}
\hskip 1in
\begin{minipage}{5cm}
\scalebox{.4}{\includegraphics{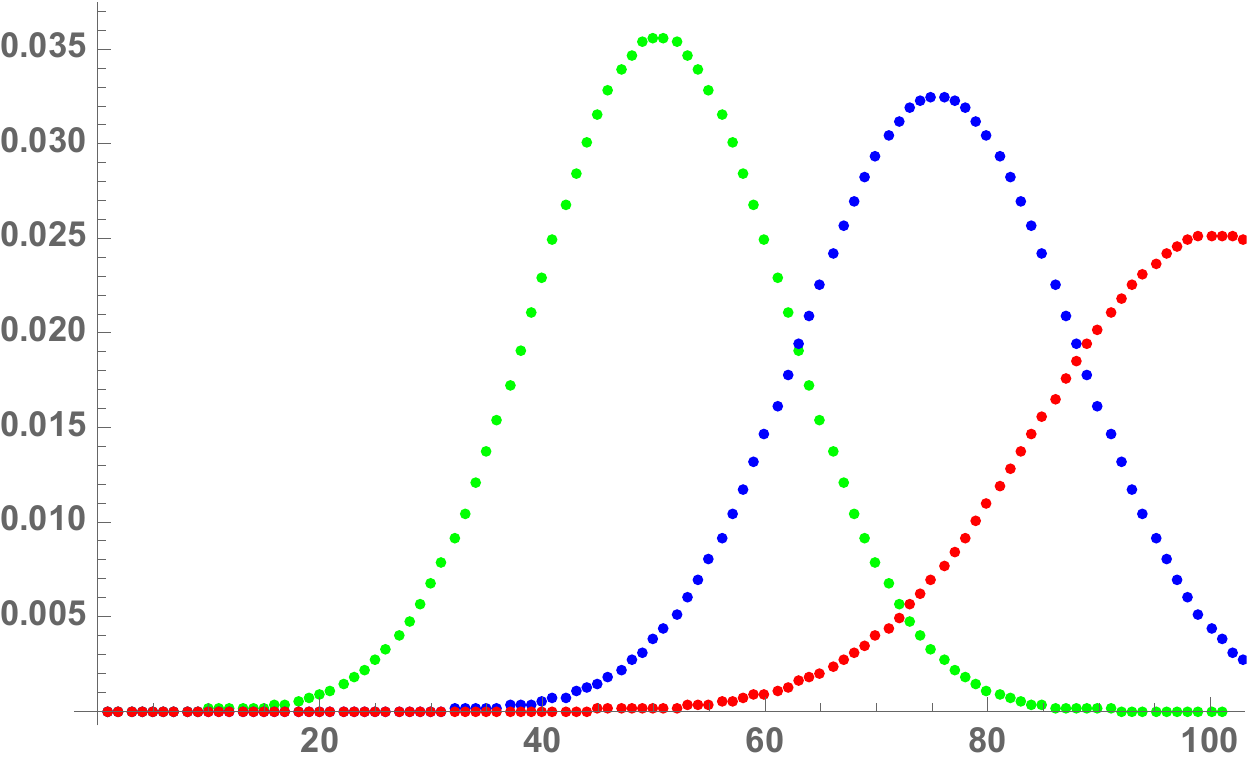} }
\caption{Values of $\lambda_s - \lambda_{s+1} $ for
$n = 100, 150, 200$, for the three curves from left to right}
\label{g2}
\end{minipage}
\end{center}
\end{figure}

The maximal difference between nearby values of $s$ will occur near the
midpoint of the transition from $1$ to zero.
For this, $s$ is also large and one can use a semiclassical or steepest descents
method. Making a change of variable to
$t = u/(1+ u)$, 
 (\ref{B1}) can be written as an incomplete beta function,
 \beqar
 \lambda_s &=& { (n+k)! \over \Gamma(s+k) (n-s)!}
 \int_0^w dt \, t^{s+k-1} (1-t)^{n-s} \nonumber\\
 &=&{ (n+k)! \over \Gamma(s+k) (n-s)!} \int_0^w \, dt\, e^{F(t)}
 \label{B7}\\
 F(t)&=&(s+k-1) \log t + (n-s) \log(1-t)\nonumber
\eeqar
where $w = R^2/(1+R^2)$. The minimum of $F(t)$ occurs
at $t_0 = (s+k -1) /(n+k -1) $. Expanding $F(t)$ around this value, 
$E^{F(t)}$ becomes a Gaussian function centered around this value of $t$.
If the maximum of the Gaussian is well within the range of integration,
which is the case for small $s$, we will find $\lambda_s \approx 1$.
If the center of the Gaussian is well beyond the range fo integration, which is the
case for $s$ near $n$, we will find
$\lambda_s \approx 0$. The midpoint of the transition occurs for
$t_0$ at the upper limit of integration, namely, for
$t_0 = w$, which corresponds to $s=s_*$ given by
\beq
s_* = w (n+k -1) - (k-1), \hskip .2in
n-s_* =  (1-w) (n+k-1) 
\label{B8}
\eeq
Expanding $F(t)$ around $t_0$, we obtain
\beqar
F(t) &=& (n+k -1) \Bigl[ \left[ t_0 \log t_0 + (1-t_0) \log (1-t_0) \right]
-   {x^2\over 2 t_0 (1-t_0) }  \nonumber\\
&&\hskip 1in +  { (1 -2 t_0) x^3\over  3 t_0^2 (1-t_0)^2}+ \cdots
\Bigr]
\label{B9}
\eeqar
where $x = t- t_0$. Using this expression, we find
\beq
\lambda_{s_*} \approx { (n+k)! \over \Gamma(s_*+k) \Gamma(n-s_* +1)}\, e^{F(t_0)}
\left[ \sqrt{\pi\, t_0 (1-t_0) \over 2 (n+k-1)} -
{2(1 -2 t_0)\over 3 (n+k-1)} + \cdots\right]
\label{B10}
\eeq
where $t_0 =w$.
It should be noted that the range of integration for $x$ for this case
is from $-\infty$ to zero.

We do a similar calculation for $\lambda_{s_* +1}$.
For $s= s_* +1$, the minimum of $F(t)$ occurs at
\beq
t_1 = {s_* +k \over n+k-1} =  w + \epsilon, 
\hskip .2in \epsilon = {1\over n+k -1}
\label{B11}
\eeq
This is just beyond the range of integration. The integral over $x$ now becomes
\beq
\int_0^w dt \, e^{F(t)} \approx e^{F(t_1)} \int_{-\infty}^{ -\epsilon} dx 
\exp\left( - {(n+k-1) x^2 \over 2 t_1 (1-t_1)}  + {(n+k-1) (1-2 t_1) x^3 \over 3 t_1^2 (1-t_1)^2 }
+\cdots \right)
\label{B12}
\eeq
The integral with the upper limit as zero is similar to what was encountered
for $\lambda_{s_*}$, but we have to subtract out the
integral from $- \epsilon$ to zero. Apart from this, the result is of the form
in (\ref{B10}) with $s_* \rightarrow s_* +1$ and
with $t_0 \rightarrow t_1= w +\epsilon$. The final result is thus
\beqar
\lambda_{s_* +1} &\approx& { (n+k)! \over \Gamma(s_*+k + 1) \Gamma(n-s_*)}\, e^{F(w+\epsilon )}
\left[ \sqrt{\pi (w+\epsilon) (1-w - \epsilon) \over 2 (n+k-1)} -
{2(1 -2 w- 2 \epsilon)\over 3 (n+k-1)} + \cdots\right]\nonumber\\
&&\hskip .2in - \epsilon \,{ (n+k)! \over \Gamma(s_*+k + 1) \Gamma(n-s_*)}\, e^{F(w+\epsilon )} 
\label{B13}
\eeqar
The rest of the simplification is straightforward, using properties of the
$\Gamma$-functions. The leading term in the difference
$\lambda_{s_*} - \lambda_{s_* +1}$ comes from the second line of
(\ref{B13}).
This leads to the expression (\ref{vN42}) quoted in text,
\beq
{\lambda_{s_*} - \lambda_{s_* +1} \over \lambda_{s_*}}
= \sqrt{2 \over w (1-w) \pi}  \, {1\over \sqrt{n+k-1}} + \cdots, \hskip .2in
\label{B14}
\eeq

The results obtained in this appendix
can also be checked numerically. 
As an example, consider the case of the two-sphere or $k = 1$.
In this case, taking $R = 1$ (i.e., $w =\half$), we have plotted, in
Fig.\,\ref{g1}, 
we have plotted the values of $\lambda_s$ for $n = 200$;
the result
shows that the values are close to zero and $1$ at the two ends and 
has the largest separation between $\lambda_s $ and $\lambda_{s+1}$ for 
$s_* = n/2$. The second graph (Fig.\,\ref{g2}) shows the
differences $(\lambda_{s_*}- \lambda_{s_*+1})$.
We see that the maximum values of the differences decrease as we increase $n$.
The peak value can be checked to
be as given by (\ref{B14}) or (\ref{vN42}).

\section*{Appendix B: Direct calculation of background field dependence}
\def\theequation{B\arabic{equation}}
\setcounter{equation}{0}

In this appendix, we will go over the explicit calculation of some of the terms which arise in the
dependence of $u^*_s u_s$ on the background fields.
In other words, we go over the calculation of the terms in
(\ref{vN52}), (\ref{vN53}). The basic strategy is to consider
$\Tr ( P A_0 )$, rather than just the the function (\ref{vN44a}) for $P$, since this can be related to the 
calculation of the effective action. 
For simplicity, we will consider only the changes in the background gauge fields,
keeping the spin connection as $\omega^{(0)}$.
$\Tr ( P A_0 )$ can be expressed as an integral
over the phase space with the star product of the functions corresponding to $P$ and
$A_0$ as the integrand. But instead of considering
$P$ and $A_0$ as defined by the modified wave functions, we can use the
wave functions for the background $(a, \omega^{(0)})$, but use
$\A$ which is a function of $A_0$ and $A_i$.
In other words,
\beqar
\Tr (P A_0) &=& \int {\widetilde{(P)} } * (A_0)\nonumber\\
&=& \int d\mu \sum_{i j} \bigl[ u^*_i P_{ij} u_j \bigr]_{a, \omega^{(0)}} * \A\nonumber\\
&=& \int d\mu \, (P)^{(0)} \, \A + {\rm terms ~with ~derivatives
~of}~ (P)^{(0)}, \, \A
\label{A1}
\eeqar
This shows that if we can identify $\A$, then from the first and third lines of this equation, we see
that we can obtain the relation between ${\widetilde{(P)} }$ and 
$(P)^{(0)}$, by functional differentiation with respect to $A_0$.
The calculation of $\A$ has been done in a few different ways; we will go over two methods.
 
The first method is essentially classical,
and can be applied to the case when the background field is Abelian \cite{VPN1}.
Let $\Omega$ denote the symplectic structure of the phase space, say, ${\mathbb{CP}^k}$;
$\Omega$ is a multiple of the K\"ahler form.  The symplectic potential
is the Abelian background gauge potential $a$, so that 
$\Omega = d a$. 
Changing the gauge field
is equivalent to using $\Omega + F = d (a +A)$ as the symplectic two-form.
We are interested in $\Tr (P A_0)$ calculated using wave functions with the background
$a+A$.  The classical version of this is the integral of $A_0$
over the phase volume corresponding to $\Omega + F$.
So we can write the equivalent of
(\ref{A1}) as
\beq
\int d\mu_{\Omega +F}\, A_0 = \int d \mu \, (P)^{(0)}  * \A =
 \int d \mu \, (P)^{(0)}  \, \A + {\rm derivative ~terms}
 \label{A2}
 \eeq
 
The two-forms $\Omega$ and $\Omega +F$ must belong to the same topological class, so that, upon
quantization, we get the same number of states for the Hilbert space.
This means that we can use a diffeomorphism to map $\Omega +F $ to $\Omega $.
We can then identify $\A$ as the image of $A_0$ under this map.
More explicitly, there is a diffeomorphism changing the local coordinates $v$ as
$v \rightarrow v-w$ such that
\beq
\Omega + F \Bigr]_{v-w} = \Omega\Bigr]_v,  \hskip .2in
\A = A_0\Bigr]_{v-w}
\label{A3}
\eeq
Equivalently, we can write
$a +A\Bigr]_{v-w} - a\Bigr]_v = d f$ for some function $f$. Taking $A$ to be a first order correction, we can solve this equation for $w$ as a series. To the quadratic order, the equations for
$w$ are
\beqar
&&w_1^j \del_j a_i  + a_j \del_i w_1^j - A_i \approx 0\nonumber\\
&&w_2^j \del_j a_i + a_j \del_i w_2^j + w_1^j \del_j A_i + A_j \del_i w_1^j
- {1\over 2} w_1^k w_1^l \del_k \del_l a_i -
w_1^k \del_k a_j \del_i w_1^j \approx 0
\label{A4}
\eeqar
where $\approx$ denotes equivalence up to an exact form. The solution to this order is
\beqar
&&w_1^j = - (\Omega^{-1})^{jk} A_k\nonumber\\
&&w_2^j = - (\Omega^{-1})^{jk} \left[
F_{kl} w_1^l + {1\over 2} w_1^m w_1^n \del_m \Omega_{nk}
+ {1\over 2} (w_1^m \del_k w_1^n) \Omega_{mn}\right]
\label{A5}
\eeqar
The expression for $\A = A_0\Bigr]_{v-w}$ obtained in this manner is
\beq
 \Omega^k \, \A =  \Omega^k \, A_0 + k\, \Omega^{k-1} \, A d A_0 + 
 {\half} {k (k-1)} \Omega^{k-2} dA A dA_0 + \cdots +
{\half}  {k} \Omega^{k-1} d (u\cdot A\, A ) + \cdots
\label{A6}
\eeq
where $ u^i = (\Omega^{-1})^{ij} \del_j A_0$. 
Using this in (\ref{A2}), we find
\beqar
{\widetilde{(u_s^* u_s)}}_{A_i \neq 0 } &=& d\mu\, (u_s^* u_s)_{A_i = 0 } 
+ {1\over N} d \left[ (u_s^* u_s)_{A_i = 0} \, Q (A, \omega)\right]\nonumber\\
&&- {k \over 2 N} d \left[ \Omega^{k-1} A (\Omega^{-1})^{ij} \del_i (u_s^* u_s)_{A=0} \, A_j
\right] +\cdots\nonumber\\
Q (A)&=& k \int_0^1 dt\, A\, ( \Omega + t\, dA )^{k-1} 
\label{A7}
\eeqar
There are several observations to be made about this expression.
First of all, $\Omega = n\, \Omega_K$ where $\Omega_K$ is the K\"ahler two-form
for the space under consideration and $N \approx n^k/k!$.
 For ${\mathbb{CP}^k}$, we have
\beq
\Omega_K = i \left[ {d z ^i ~d\bz^i \over (1+\bz\cdot z)}
- { \bz \cdot dz ~ z\cdot d\bz \over (1+\bz\cdot z )^2}
\right]
\label{A8}
\eeq
Thus terms with $\Omega^{k-1}$ are down by a power of $n$ compared to $\Omega^k$-terms.
Also, $(\Omega^{-1})^{ij}$ gives an additional power of $1/n$. 
The series represented by $Q(A, \omega )$ has $(k-1)$ terms terminating
with $\Omega^0$. 
Integrating (\ref{A8}) over the region $D$, we find
\beq
\lambda_s\Bigr]_{A_i \neq 0} - \lambda_s\Bigr]_{A_i = 0} =
{Q (A) \over N} \, (u_s^* u_s)_{{\rm at}~R} - {k\over 2 N} \left[ \Omega^{k-1} \, (\Omega^{-1})^{ij} \del_i (u_s^* u_s)\, A_j \right]_{\del D} + \cdots
\label{A9}
\eeq
Comparing this with (\ref{vN53}), we see that we can identify
\beq
W^i \bigl[\del_i(u^*_s u_s) \bigr]_{\del D; a, \omega^{(0)}} 
=  - {k\over 2 N} \left[ \Omega^{k-1} \, (\Omega^{-1})^{ij} \del_i (u_s^* u_s)\, A_j \right]_{\del D}
\label{A10}
\eeq
This term is order $A /n^2$ while the leading term involving $Q$ is
of order $A/n$.

The second method is to consider the time-evolution of the occupancy matrix \cite{Kar1}.
This should be a unitary transformation of the form
$U(t) P_0\, U^\dagger (t)$. The action governing the time-evolution can then be written as
\beq
\S = \int dt \, \Tr \left[ P_0 \left( U^\dagger \, i {\del U \over \del t} - U^\dagger \, \A\, U \right)
\right]
\label{A11}
\eeq
It is easy to verify that the variational equation for this is the (quantum) Liouville equation for
$P = U(t) P_0 U^\dagger (t)$. The action (\ref{A11}) can be rewritten using star products as
\beq
\S = N  \int d\mu ~dt~ \left[ i ({P_0} *{  U}^\dagger  * \del_t { U})
~-~ ({ P_0} *{U}^\dagger  * { \A} * {U}) \right]  
\label{A12}
\eeq
where $N$ is again the degeneracy. 
In (\ref{A12}), all quantities are $c$-number functions, the symbols which use wave functions
defined with the background fields $A_i$.
Rather than working out the perturbed wave functions and symbols directly, we note that the action has the gauge symmetry
\beq
\delta U = - i \theta * U  ,\hskip .5in~
\delta \A (\vec{x}, t)  = \del _t \theta (\vec{x}, t)\, -\!  i \left( \theta* \A \!-\!  \A  * \theta \right)
\label{A13}
\eeq
for some function $\theta$ on $\M$.
The background gauge fields are only defined up to the gauge symmetry
\beq
\delta A_\mu = \del_\mu \Lambda + i [ a_\mu + A_\mu , \Lambda ], \hskip .2in 
\delta a_\mu = 0
\label{A14}
\eeq
for some function $\Lambda$ on $\M$.
Since (\ref{A13}) is the only gauge symmetry for the action (\ref{A12}), the transformation
(\ref{A14}) must induce a transformation of the form
(\ref{A13}). Thus we must have $\A$ and $\theta$ as functions of $a_\mu, \, A_\mu$, $\Lambda$
such that $
\A_\mu (A_\mu + \delta A_\mu )
\approx \A + \delta \A$ with $\delta A_\mu$ as in
(\ref{A14}) and $\delta \A$ as in (\ref{A13}).
Taking $\A$ to be $A_0$ to the lowest order, we can use this idea to solve for
$\A$ in terms of the star product (defined in terms of $\Omega$). The field $U$ is a boundary field at the edge of the occupied states and can be set to
the identity at the end of the calculation. This strategy was used in \cite{Kar1} and
leads to
\beq
\A = A_0 + {1\over 4} (\Omega^{-1})^{ij} \{ A_i,
2 D_j A_0 + i [A_j, A_0] \} + \cdots
\label{vN63}
\eeq
This result is identical to the previous one, if the fields are Abelian. 

 \bigskip
I thank Dimitra Karabali and
Alexios Polychronakos for reading the
manuscript carefully and providing
many useful comments which have significantly improved the presentation.
I also thank A. Polychronakos for pointing
a missing term in an earlier derivation of the formula (\ref{vN42}).
I am also grateful to
A. Abanov for
comments and to A. Abanov and P. Ghaemi for some of the relevant references.

This research was supported in part by the U.S.\ National Science
Foundation grant PHY-1820721
and by PSC-CUNY awards.



\begin{thebibliography}{99}
\bibitem{TJ} The recognition of the relation between entropy and gravity goes back to the work of
J.Bekenstein and S. Hawking from the 1970s, but for the more modern approach discussing the Einstein equations as a  thermodynamic relation, see
T. Jacobson, \PRL~{\bf 75}, 1260 (1995); \PRL~{\bf 116}, 201101 (2016);
T. Padmanabhan, Rep. Progr. Phys. {\bf 73}, 6901 (2010).

\bibitem{Ryu-Tak} There are many papers on entropy in the holographic framework 
starting with S. Ryu and T. Takayanagi, \PRL~{\bf 96}, 181602 (2006);
\JHEP~0608:045 (2006).

\bibitem{raamsdonk} T. Faulkner {\it et al}, arXiv:1312.7856;
 B. Swingle and M. van Raamsdonk, \\
 arXiv:1405.2933.
 
\bibitem{verlinde} E. Verlinde, \JHEP 1104:029 (2011).

\bibitem{CS} A. Connes and E. Stormer, J. Funct. Analysis, {\bf 28}, 187 (1978).

\bibitem{vN} R. Clifton and H. Halverson, Stud. Hist. Philos. Mod. Phys. {\bf 32}, 1 (2001);
S. Hollands and K. Sanders, arXiv:1702.04924[quant-ph];
C.J. Fewster and K. Rejzner, arXiv:1904.04051.

\bibitem{HM} H. Halverson and M. Mueger, arXiv:math-ph/0602036.

\bibitem{witten} E. Witten \RMP~{\bf 90}, 045003 (2018) (arXiv: 1803.04993[hep-th]).

\bibitem{NC} A. Connes, {\it
Nocommutative Geometry}  (Academic Press, 1994); J.
Madore, {\it An Introduction to Noncommutative Geometry and its Physical
Applications}, LMS Lecture Notes 206 (1995); G. Landi,
{\it An Introduction to Noncommutative Spaces and their
Geometry}, Lecture
Notes in Physics, Monographs m51 (Springer-Verlag, 1997);
For another recent review of fuzzy spaces and theories defined on them, see, 
A.P. Balachandran, {\it Pramana} {\bf 59} (2002) 359;
A.P. Balachandran and S. Kurkcuoglu, \IJMP ~{\bf A19} (2004) 3395;
A.P. Balachandran, S. Kurkcuoglu and S. Vaidya, hep-th/0511114.

\bibitem{KNrev} For a discussion of noncommutative spaces
using the lowest Landau level approach, see
D. Karabali and V.P. Nair, J. Phys. A Math. Gen. {\bf 39}, 12735 (2006);
D. Karabali, V.P. Nair and R. Randjbar-Daemi,
in {\it From  Fields to Strings: Circumnavigating Theoretical Physics},
Ian Kogan Memorial Collection, M. Shifman, A. Vainshtein and J. Wheater (eds.),
World Scientific, 2004; p. 831-876 and references therein.

\bibitem{KN1} D. Karabali and V.P. Nair, \NP~{ \bf B641}, 533 (2002);
\NP~{ \bf B679}, 427 (2004);
\NP~ { \bf 697}, 513 (2004).
For earlier work on Hall effect in higher dimensions, see
S.C. Zhang and J.P. Hu,  {Science}, {\bf 294} (2001) 823; 
J.P. Hu and S.C. Zhang, \PR~{\bf B66}, 125301 (2002).

\bibitem{3dgrav} A. Ach\'ucarro and P. Townsend, 
Phys. Lett. {\bf B180}, 89 (1986);
E. Witten, Nucl. Phys. {\bf B311}, 46 (1988).

 \bibitem{zanelli} For a recent general review, see
J. Zanelli, arXiv:0502193[hep-th].

\bibitem{dubail} J. Dubail, N. Read and E.H. Rezayi, \PR~{\bf B85}, 115321 (2012);
\PR~{\bf B86}, 245310 (2012).

\bibitem{rodr} I.D. Rodriguez and G. Sierra, \PR~{\bf B80}, 15303 (2009).

\bibitem{other1} H.Li and F. D. Haldane, \PRL~{\bf 101}, 010504 (2008);
Z. Liu and R.N. Bhatt, \PRL~{\bf 117}, 206801 (2016).

\bibitem{other2} For discussions fo entanglement entropy for Hall systems from a holographic point of view, see
M. Fujita, W. Li, S. Ryu and T. Takayanagi,
JHEP 06(2009) 066;
T. Takayanagi, J. Phys. Conf. Ser. {\bf 462}, 012053
(2013).

\bibitem{KN-Dolb} D. Karabali and
V.P. Nair, \PR~{\bf D94}, 024022 (2016);
\PR~{\bf D94}, 064057 (2016).

\bibitem{EGH} See, for example, T. Eguchi, P.B. Gilkey and A.J. Hanson,
Phys. Rep. {\bf 66}, 213 (1980).

\bibitem{MSZ} J. Manes, R. Stora and B. Zumino, \CMP~{\bf 102}, 157 (1985).

\bibitem{ray-sakita} R. Ray and B. Sakita, Ann. Phys. {\bf 230}, 131
(1994); \PR~{\bf B65}, 035320 (2001).

\bibitem{AS} See, for example,
M. Abramowitz and I.A. Stegun, {\it Handbook of Mathematical
Functions}, Dover Publications, New York (1965), p. 263.

\bibitem{VPN1} V.P. Nair, \NP~ {\bf B750}, 289 (2006).

\bibitem{Kar1} D. Karabali,  \NP~{ \bf B726}, 407 (2005); \NP~ { \bf B750}, 265 (2006).


\end{thebibliography}
\end{document}